\documentclass[epj]{svjour}
\usepackage{graphics}
\begin{document}
\title{High pressure effects on the benzene pre-crystallization metastable states}
\author{Mustapha Azreg-A\"{\i}nou\inst{1} \and Beycan \.{I}brahimo\v{g}lu\inst{2}
}                     
\offprints{}          
\institute{Ba\c{s}kent University, Engineering Faculty, Ba\u{g}l\i ca Campus, Ankara, Turkey \and Anatolian Plasma Technology Energy Center, Gazi University G\"{o}lba\c{s}i Campus, Ankara, Turkey}
\date{Received: date / Revised version: date}
%
\abstract{
We report new results on the liquid to solid phase transition of benzene. We determine experimentally and investigate the properties of a number of parameters of the benzene metastable state under different pressures (from 0.1 up to 2200 atm). It is shown that the supercooling, pressure drop, incubation period, time of abrupt transition from the metastable state to the crystalline state, and time of isothermal freezing all decrease as the external pressure increases, then they all vanish at 2200 atm and 356 K which may mark the end-point of metastability. Quadratic interpolation formulas for these parameters are provided. The densities and molar heat capacities of supercooled benzene under different pressures have been calculated too.
\PACS{
     {}{}   \and
     {}{}
     } 
} 
\maketitle
\section{Introduction}
\label{seci}
In modern science the investigation of the metastable states of matter has always been a fruitful and a fascinating subject. These are relatively stable states with an energy higher than that of the state of the system with least energy (the ground state). They have a longer lifetime than the other excited states and, in contrast, a shorter lifetime than the state of least energy. Applications of metastability results range from atomic spectroscopy (energy levels of atoms and light emission), population inversion in lasing media (the population of the metastable state can exceed the population at the ground state), to thermodynamics (for instance, diamond, polymers, martensite, and fullerite are manifestations of metastability).

In thermodynamics metastable phases behave just like stable states; in that, within a certain region in the temperature pressure plane~\cite{Braz}, the metastable phases, if subject to undergo phase transitions, the latter are reversible and obey the equilibrium laws of thermodynamics.

Benzene is a prototype of the aromatic hydrocarbons and it is an object of numerous experimental and theoretical studies~\cite{1,2,3,4,5,8,6,25,26} (other references will be cited sequentially).  In the literature, the benzene pre-crystallization metastable state at normal pressure has been mainly studied by the thermal pre-treatment of the liquid phase; this pre-treatment influences the degree of supercooling (undercooling)  relative to the melting temperature  $T_{\rm fus}$~\cite{6,7}. Moreover, to the best of our knowledge, there are practically no studies of the pressure effect on the liquid benzene pre-crystallization metastable state. A first investigation of this kind has been carried out in Ref.~\cite{8} where we have followed a technique similar to the one presented in this work.  We have been able to determine the values of the molar volume and enthalpy changes at the solidification points under permanent high external pressures ranging from 20.6 to 102.9 MPa.

In the present work we studied experimentally the effect of pressure on the behavior of benzene metastable states.

\section{Materials and Methods\label{secet}}
Benzene was purchased from Cherepovet Nitrogen Factory and used without further purification~(Table~\ref{Tab0}).
\begin{table}[ht]
	\caption{Material investigated: origin and purity}
	\label{Tab0}
		\begin{tabular}{cccc}
			\hline\noalign{\smallskip}
			\textbf{chemical}   & \textbf{source}    & \textbf{initial mass}  & \textbf{purification} \\
			\textbf{name}   &    & \textbf{fraction purity}   & \textbf{method}  \\
			\noalign{\smallskip}\hline\noalign{\smallskip}
			Benzene  & Cherepovet & 0.998 & none  \\
			& Nitrogen &  &   \\
			& Factory &  &  \\
			\noalign{\smallskip}\hline\noalign{\smallskip}
	\end{tabular}
\end{table}

\begin{figure}
	\resizebox{0.49\textwidth}{!}{%
		\includegraphics{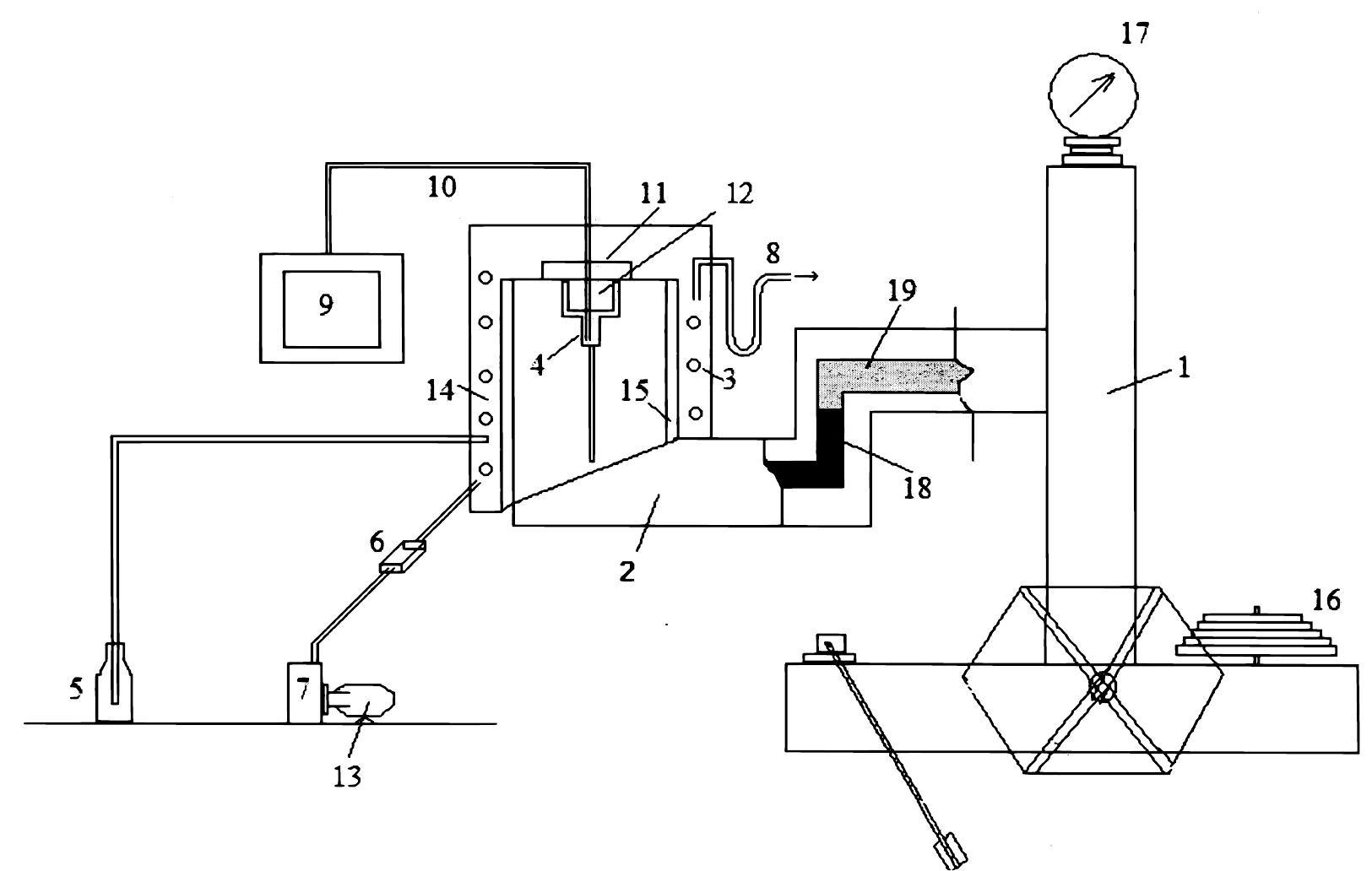}
	}
	\caption{Experimental facility diagram: 1: manometer, MP-2500, 2: compressor, 3: cooling agent,
		4: measuring container, 5: Dewar container for liquid  nitrogen, 6: regulator, 7: vacuum pump, 8: flow-meter, 9: potentiometer KSP-4, 10: thermocouples, 11: serpentine (capillary tube of a coil), 12: container, 13: engine, 14: copper tubes, 15: copper pipes, 16: counterweight, 17: manometer, 18: mercury, 19: lubricant.}
	\label{Fig1}       
\end{figure}

With the help of the experimental facility~shown in Fig.~\ref{Fig1} we have measured and controlled the following parameters of the benzene metastable state. The pressure $p$, the container temperature, the freezing (crystallization or solidification) temperature  $T_{\rm cr}$, the temperature $T_{\rm n}$  in the low point of the metastable state (i.e. the nucleation temperature that we could reach by cooling the liquid benzene before it turned into a solid state), the supercooling  $\Delta T\equiv T_{\rm cr}-T_{\rm n}$, the pressure drop or difference $\Delta p$  at the initial stage of explosive crystallization, the time $t_1$ is the incubation period of the liquid phase in the metastable state, the time $t_2$ of abrupt transition from the metastable state to the crystalline state, the time $t_3$ of isothermal freezing, and the total time  of solidification $t_{tot}=t_1+t_2+t_3$. Referring to Fig.~\ref{Fig2}, $\Delta T$ is the temperature at point $b$ minus the temperature at point $c$ and $\Delta p$ is the pressure at point $c$ minus the pressure at point $d$.

To carry out the experiments, the whole system was vacuumized. The measuring vessel (4) was filled with benzene of volume 10 cm$^3$. Before cooling, the whole system was correspondingly pressurized. It was evident from the experiments that the system cooling rate and pressure had a direct influence on the metastable state of benzene. To maintain the same 0.85 K/s cooling rate in all experiments, we controlled the flow of the nitrogen vapor delivery from the Dewar container (5) into the copper tubes (14) with the help of the vacuum pump (7). The rate of nitrogen vapor feeding was regulated by the regulator (6). The system was pressurized with the help of the pneumatic press MP-2500 (1):  It is the most accurate and stable pressure device that gives the best repeatability of the measurement results. The technical specifications of the manometers are compatible with GOST R 8802-2012. The pressure was measured by a dead-weight pressure gauge tester with a relative standard uncertainty $u_{\rm r}(p)=0.005$. The temperature was measured by the self-potentiometer KSP-4 (9) and the Chromel Copel thermocouple. The thermocouple was kept straight in the benzene sample. The standard uncertainty of KSP-4 devices is $u(T)=0.2$ K. The thermo-grams were recorded in the temperature-time,  $T$-$t$, coordinates with the pressure being fixed. All measures were performed in the range of temperatures from 278.5 K up to 368.0 K.

\section{Experimental results\label{secer}}
\begin{figure*}
	\centering
	\resizebox{0.65\textwidth}{!}{%
		\includegraphics{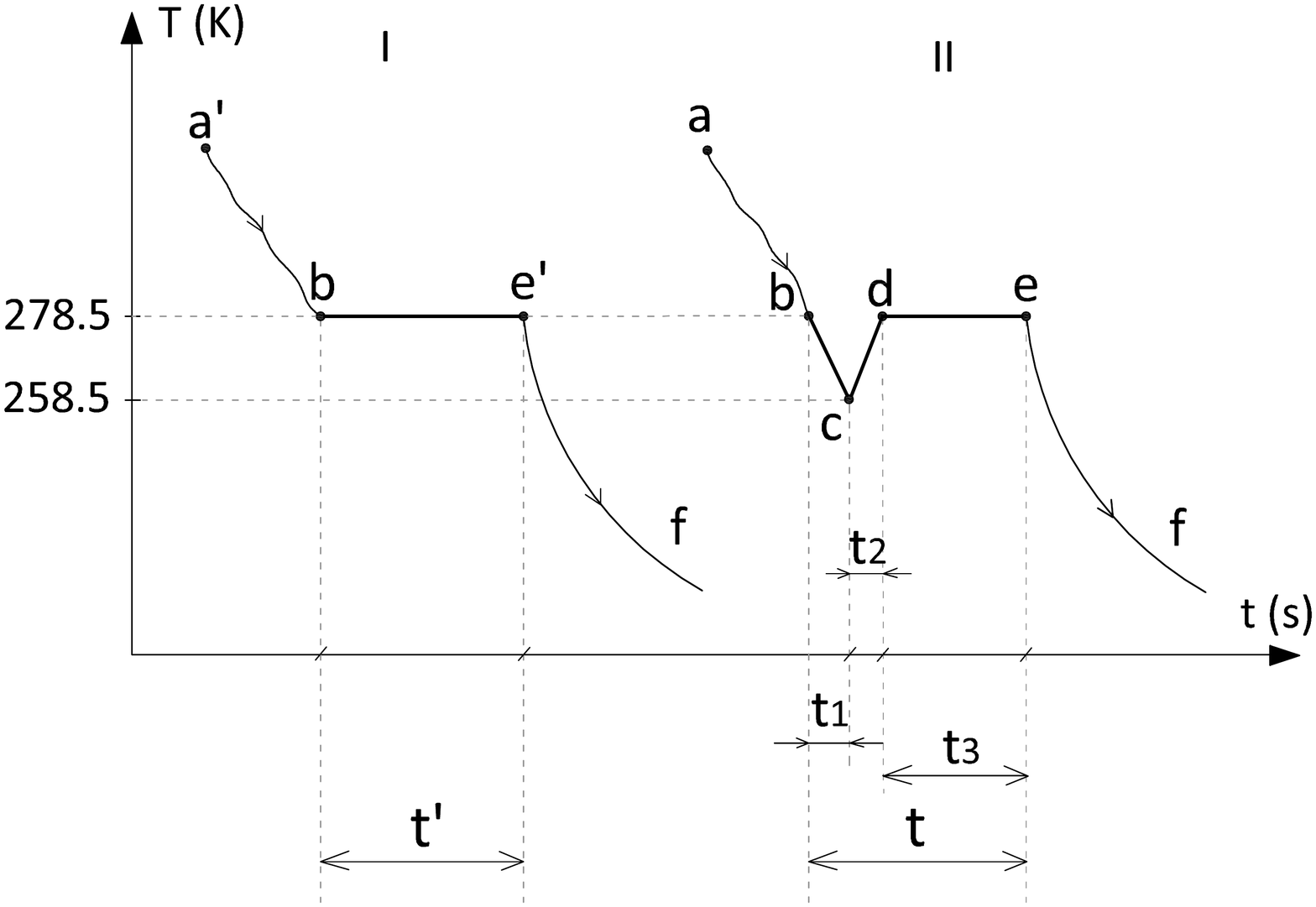}} \\
	\caption{Schematic thermo-grams in the $T$-$t$ coordinates recorded at $p = 0.1$ atm. They characterize (I) the absence of a metastable state and equilibrium crystallization and (II) the availability of a metastable state and non-equilibrium-explosive crystallization. The supercooling  $\Delta T\equiv T_{\rm cr}-T_{\rm n}$ is the temperature at point $b$ minus the temperature at point $c$ and the pressure drop $\Delta p$ at the initial stage of explosive crystallization is the pressure at point $c$ minus the pressure at point $d$.}\label{Fig2}
\end{figure*}

Let us first analyze two schematic thermo-grams of benzene cooling (Fig.~\ref{Fig2}) without (I) and with (II) a metastable area of benzene of volume 10 cm$^3$ at 0.1 atm.
\begin{table*}
	\centering
	\caption{Parameters of the metastable and freezing states of benzene: the external pressure$^a$ $p$, the freezing temperature $T_{\rm cr}$ ($=T_d=T_e$ Fig.~\ref{Fig2}), the temperature $T_{\rm n}$  in the low point of the metastable state (i.e. the nucleation temperature that we could reach by cooling the liquid benzene before it turned into a solid state), the supercooling  $\Delta T\equiv T_{\rm cr}-T_{\rm n}$, the pressure drop $\Delta p$  at the initial stage of explosive crystallization, the incubation period $t_1$  of the liquid benzene stay in the metastable state, the time $t_2$ of an abrupt transition from the metastable state to the crystalline state, the time $t_3$ of isothermal freezing, and the total time  of solidification $t_{tot}=t_1+t_2+t_3$. Referring to Fig.~\ref{Fig2}, the supercooling  $\Delta T\equiv T_{\rm cr}-T_{\rm n}$ is the temperature at point $b$ minus the temperature at point $c$ and the pressure drop $\Delta p$ at the initial stage of explosive crystallization is the pressure at point $c$ minus the pressure at point $d$.}
	\label{Tab1}
	\begin{tabular}{c|c|c|c|c|c|c|c|c|c|c}
		\hline\noalign{\smallskip}
		$p$   & $T_{\rm cr}$    & $T_{\rm n}$   & $\Delta T$ & $p_c$   & $p_d$ & $\Delta p$ & $t_1$  &$t_2$   & $t_3$   &  $t_{tot}$   \\
		atm   & K    & K   & K & atm  & atm  & atm  & s  & s & s & s \\
		\hline\noalign{\smallskip}
		0.1  & 278.5 & 258.5 & 20.0       & ---    & ---      & ---  & 160.0 & 6.0 & 260.0 & 426.0 \\
		100.0  & 279.0 & 267.0 & 13.0     & 100.0  & 68.0     & 32.0 & 131.0 & 3.0 & 210.0 & 343.0 \\
		200.0  & 280.0 & 270.0 & 10.0     & 200.0  & 173.0    & 27.0 & 112.0 & 2.5 & 180.0 & 294.0 \\
		300.0 & 282.8 & 274.6 & 8.2       & 300.0  & 277.0    & 23.0 & 97.0  & 2.0 & 155.0 & 254.0 \\
		400.0  & 284.8 & 277.8 & 7.0      & 400.0  & 380.0    & 20.0 & 85.0  & 2.0 & 135.0 & 222.0 \\
		500.0  & 286.7 & 280.7 & 6.0      & 500.0  & 482.5    & 17.5 & 72.0  & 2.0 & 115.0 & 189.0 \\
		600.0  & 289.5 & 284.5 & 5.0      & 600.0  & 585.0    & 15.0 & 61.0  & 1.5 & 97.0  & 159.0 \\
		700.0  & 291.2 & 287.2 & 4.0      & 700.0  & 687.0    & 13.0 & 53.0  & 1.5 & 85.0  & 139.0 \\
		800.0  & 294.8 & 291.2 & 3.6      & 800.0  & 788.5    & 11.5 & 44.0  & 1.5 & 70.0  & 115.0 \\
		900.0  & 297.0 & 293.8 & 3.3      & 900.0  & 890.5    & 9.5  & 37.0  & 1.5 & 60.0  & 98.0  \\
		1000.0 & 299.5 & 296.6 & 2.9      & 1000.0 & 992.0    & 8.0  & 31.0  & 1.0 & 50.0  & 82.0  \\
		1100.0 & 302.5 & 300.1 & 2.4      & 1100.0 & 1093.3   & 6.7  & 25.0  & 1.0 & 40.0  & 66.0  \\
		1200.0 & 305.3 & 303.3 & 2.0      & 1200.0 & 1194.5   & 5.5  & 22.0  & 1.0 & 35.0  & 58.0  \\
		1300.0 & 308.0 & 306.2 & 1.8      & 1300.0 & 1295.8   & 4.2  & 18.0  & 1.0 & 30.0  & 49.0  \\
		1400.0 & 312.2 & 310.7 & 1.5      & 1400.0 & 1396.4   & 3.6  & 16.0  & 1.0 & 25.0  & 42.0 \\
		1500.0 & 315.5 & 314.3 & 1.1      & 1500.0 & 1497.2   & 2.8  & 12.0  & 1.0 & 19.0  & 32.0  \\
		1600.0 & 320.1 & 319.2 & 0.9      & 1600.0 & 1598.1   & 1.9  & 8.0   & 1.0 & 14.0  & 23.0  \\
		1700.0 & 324.0 & 323.3 & 0.7      & 1700.0 & 1698.8   & 1.2  & 6.0   & 0.5 & 10.0  & 16.0  \\
		1800.0 & 328.0 & 327.5 & 0.5      & 1800.0 & 1799.4   & 0.6  & 4.0   & 0.5 & 6.0   & 10.0  \\
		1900.0 & 333.5 & 333.2 & 0.3      & 1900.0 & 1899.8   & 0.2  & 2.0   & 0.5 & 4.0   & 6.0   \\
		2000.0 & 340.5 & 340.4 & 0.1      & 2000.0 & 1999.9   & 0.1  & 1.0   & 0.5 & 2.0   & 3.0   \\
		2100.0 & 347.0 & 347.0 & 0.0      & 2100.0 & 2100.0   & 0.0  & 0.5   & 0.0 & 1.0   & 1.0   \\
		2200.0 & 356.0 & 356.0 & 0.0      & 2200.0 & 2200.0   & 0.0  & 0.0   & 0.0 & 0.0   & 0.0   \\
		\hline\noalign{\smallskip}
	\end{tabular}\\
	\footnotesize{$^a$ The uncertainties are $u_{\rm r}(p) = 0.005$ and $u(T)=0.2$ K.}
\end{table*}
\begin{figure*}
	\centering
	\resizebox{0.65\textwidth}{!}{%
		\includegraphics{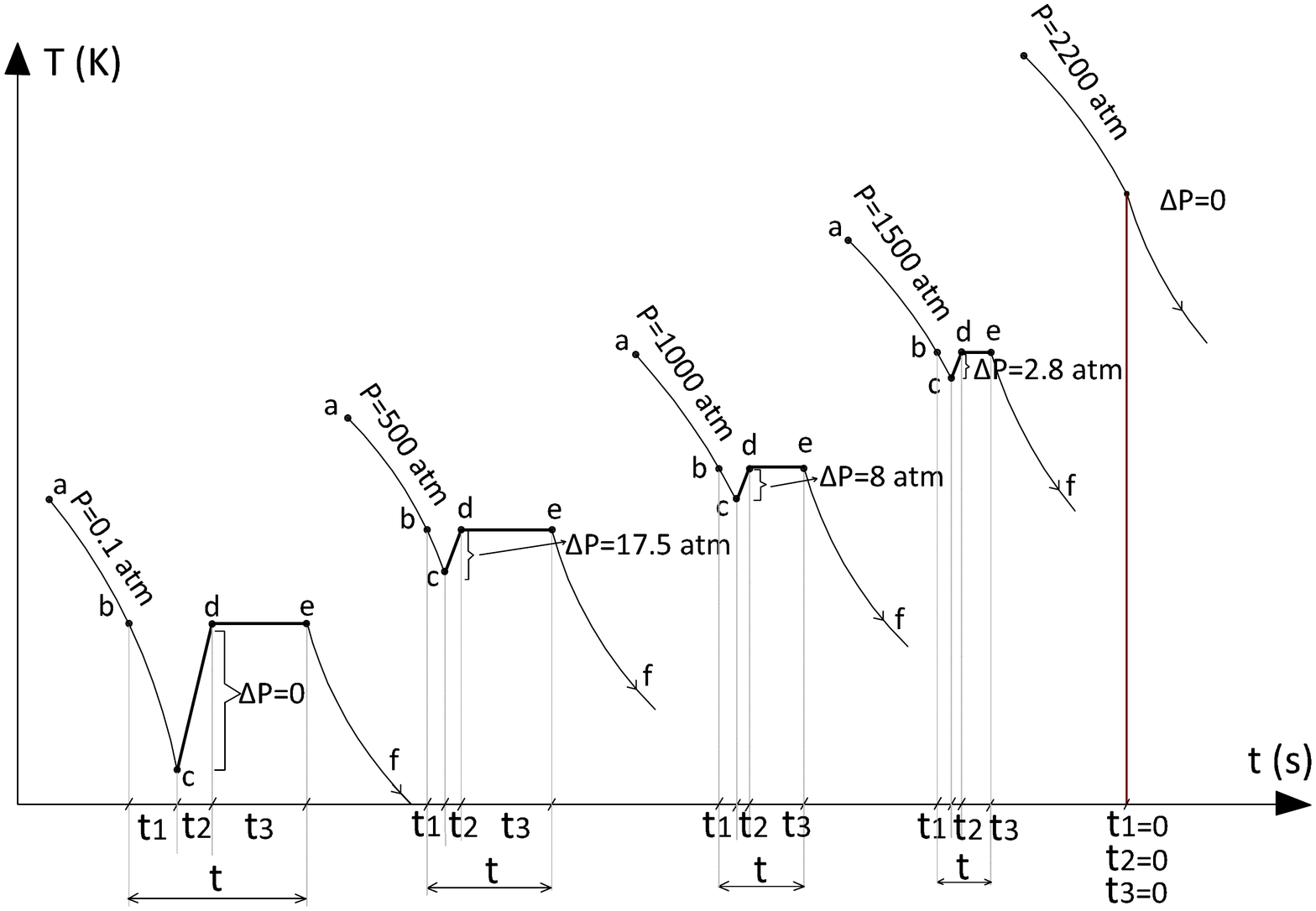}
	}
	\caption{Schematic thermo-grams recorded at pressures $p  =$ 0.1; 500; 1000; 1500 and 2200 atm. The supercooling $\Delta T$ and the pressure drop $\Delta p$ are shown on the boundaries of the metastable state.}\label{Fig3}
\end{figure*}

The first thermo-gram characterizes a supercooling-free equilibrium crystallization ($\Delta T\simeq 0$). Such thermo-grams are determined after a short phase of pre-heating of liquid benzene followed by a cooling process~\cite{6}. On the path  $a'\to b$, the liquid phase cools down and on the path $b\to e'$  the isothermal crystallization takes place at 278.5 K. This temperature coincides with the benzene melting temperature  $T_{\rm fus}$~\cite{3}. On the path  $e'\to f$, the solid benzene cools down. This corresponds to an supercooling  $\Delta T= 0$ K

If the cooling starts from the point  $a$ in the thermo-gram II (with higher temperature than that of the point $a'$ in the thermo-gram I, corresponding to an overheating of about 5 K relative to  $T_{\rm fus}$: we mostly worked with $T_{a'}=T_{\rm fus}+5$ and $T_{a}=T_{\rm fus}+15$), another shape of the $T$-$t$ curve is observed. The temperature approaches the area of supercooled states along the path  $b\to c$ with $T_{\rm n}=258.5$ K (corresponding to an supercooling  $\Delta T= 20$ K). The time $t_1\simeq 160$ s is the incubation period of the liquid phase in the metastable state. As the time $t_1$  elapses, the temperature starts rising quickly from the point $c$ to the point $d$ and the process lasts $t_2\simeq 6$ s during which the rate of adiabatic process on the segment $cd$ remains 4 K/s. Keeping in mind that the system cooling rate is $\sim$ 0.85 K/s $\ll$ 4 K/s, the heat losses into the environment can be neglected and the equation of heat balance can be written as
\begin{equation}\label{er1}
	m_x\Delta H_{\rm fus}\simeq C_p m\Delta T,
\end{equation}
where $m_x$  is the mass of the solidified part of the sample after the metastable state has elapsed, $m = 8.8$ g is the mass of the whole sample of benzene, $C_p=1.759$ kJ/(kg K) [or $C_p=0.137$ kJ/(mol K)]  is the molar heat capacity at constant pressure,  $\Delta H_{\rm fus}=128$ kJ/kg [or $\Delta H_{\rm fus}=9.98$ kJ/mol] is the enthalpy of benzene melting~\cite{9}. These values of the enthalpy and heat capacity are temperature dependent~\cite{8} and here we are providing their values at $T = 278.5$ K. From this formula one can calculate the fraction of the volume (or mass) of the solidified benzene that has reached the metastable state: $\alpha\equiv v_x/v=m_x/m=C_p \Delta T/\Delta H_{\rm fus}$ . That is $m_x\simeq 2.4$ g. Hence, in the thermo-gram II, the remaining part of benzene,  $\beta=1-\alpha=0.73$ (or 6.4 g), freezes in a time $t_3=260$ s at the temperature 278.5 K. Thus, the total time $t_{tot}$ of the whole process of solidification is 426 s.

The relative \emph{real} concentration $\alpha(t)\equiv v(t)/v$, which is a function of time, is given by Avrami equation~\cite{6}, $\alpha(t)=1-\exp(-Kt^n)$, where $K$ and $n$ are constants, is such that the remaining fraction of the liquid phase(s) $\beta(t)=1-\alpha(t)=\exp(-Kt^n)$ is a decreasing exponential function of $t$. Here $v(t)$ is the instantaneous volume of the crystallized or solidified clusters at $t$ in a metastable liquid phase. Since $t$ in bounded from above by the incubation period $t_1$ ($t\leq t_1$), in practice this concentration remains in the range of $0.37\pm 0.01$ at the end of the metastable state for all pressures.

The above balance formula~(\ref{er1}) allows one to determine an upper limit for the maximum supercooling $\Delta T_{\rm max}$. Since $m_x\ll m$, one obtains $\Delta T_{\rm max}\ll \Delta H_{\rm fus}/C_p$. For benzene this yields $\Delta T_{\rm max}\ll 73$ K. To the best of our knowledge, as table~\ref{Tab1} confirms it, this upper limit was never exceeded.

As is well known, the supercooling $\Delta T$ depends on many factors. In Sec.~\ref{sectr}, we will illustrate by example how to use the graphical method to determine an extrapolated value of the maximum supercooling, which as we shall see, is well below the upper limit of 73 K.

thermo-grams, similar to the thermo-gram II depicted in Fig.~\ref{Fig2}, have also been obtained for other static pressures $p$  up to 2200 atm. Table~\ref{Tab1} provides the temperatures $T_{\rm cr}$, $T_{\rm n}$, $T_d$, $T_e$   corresponding respectively to the points $b$, $c$, $d$, $e$ in the thermo-grams, the pressures $p_c$  and $p_d$  corresponding respectively to the points $c$ and $d$, the supercooling  $\Delta T$, the pressure drop $\Delta p$  at the transition temperature from the point $c$ up to the point $d$, and the time intervals  $t_1$,  $t_2$,  $t_3$,  $t_{tot}$ for 23 different values of the external pressure.

From table~\ref{Tab1} it is obvious that as the pressure increases the temperature of freezing $T_{\rm cr}$ increases too and, as expected, the other parameters ($\Delta T$, $\Delta p$,  and the time intervals $t_1$,  $t_2$,  $t_3$,  $t_{tot}$) decrease.

To illustrate these changes, we have depicted in Fig.~\ref{Fig3} separate thermo-grams at the pressures $p  =$ 0.1; 500; 1000; 1500 and 2200 atm. At the corresponding freezing points one can expand the functions $T_{\rm cr}=f(p)$, $T_{\rm n}=f(p)$, $\Delta T=f(p)$, $\Delta p=f(p)$, $t_1=f(p)$, and $t_{tot}=f(p)$ as follows.
\begin{equation}\label{er2}
	T_{\rm cr}=A_1+B_1p+C_1p^2,
\end{equation}
where $A_1=278.5$ K, $B_1=7\cdot 10^{-3}$ K atm$^{-1}$, and $C_1=1.235\cdot 10^{-5}$ K atm$^{-2}$.
\begin{equation}\label{er3}
	T_{\rm n}=A_2+B_2p+C_2p^2,
\end{equation}
where $A_2=258.5$ K, $B_2=3.3\cdot 10^{-2}$ K atm$^{-1}$, and $C_2=4.296\cdot 10^{-6}$ K atm$^{-2}$.
\begin{equation}\label{er4}
	\Delta T=T_{\rm cr}(p)-T_{\rm n}(p),\qquad \Delta p=A_3-B_3p+C_3p^2,
\end{equation}
where $A_3=32$ atm, $B_3=3.2\cdot 10^{-2}$, and $C_3=7.794\cdot 10^{-6}$ atm$^{-1}$.
\begin{equation}\label{er6}
	t_1=A_4-B_4p+C_4p^2,
\end{equation}
where $A_4=160$ s, $B_4=0.179$ s atm$^{-1}$, and $C_4=4.980\cdot 10^{-5}$ s atm$^{-2}$.
\begin{equation}\label{er7}
	t_{tot}=A_5-B_5p+C_5p^2,
\end{equation}
where $A_5=426$ s, $B_5=0.480$ s atm$^{-1}$, and $C_5=1.342\cdot 10^{-4}$ s atm$^{-2}$.

\begin{figure}
	\resizebox{0.49\textwidth}{!}{%
		\includegraphics{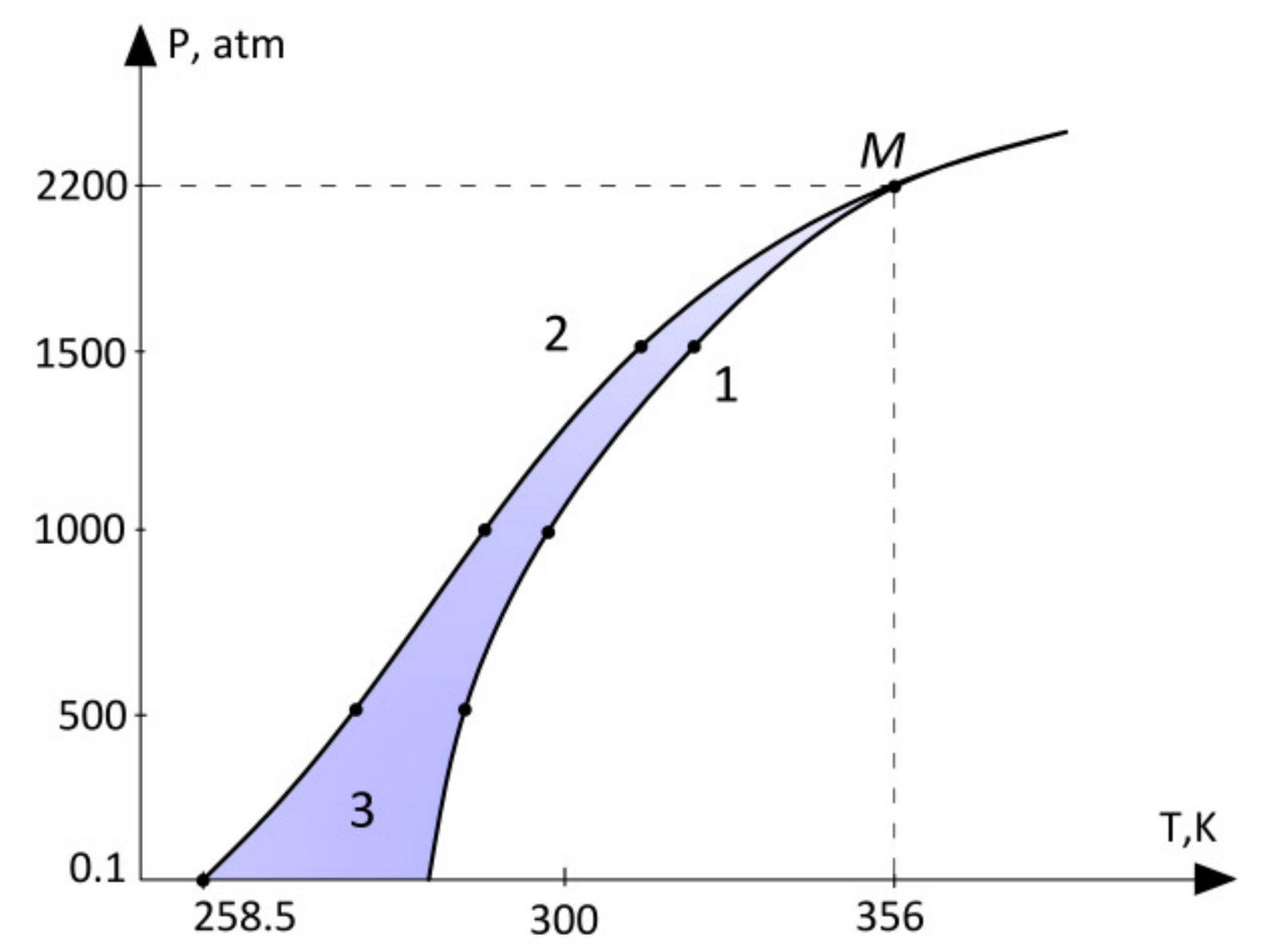}
	}
	\caption{Interpolation curves. Right curve (denoted by 1): $T_{\rm cr}=f(p)$. Left curve (denoted by 2): $T_{\rm n}=f(p)$. Dashed Area (denoted by 3): $\Delta T=f(p)$.}\label{Fig4}
\end{figure}

Relying on these data one concludes that the curves $T_{\rm cr}=f(p)$ and $T_{\rm n}=f(p)$  intersect at some point $M$  (Fig.~\ref{Fig4}) at and beyond which (that is for $T\geq 356$ K) the metastable state parameters $\Delta T$, $\Delta p$, $t_1$,  $t_2$,  $t_3$,  and $t_{tot}$ vanish.
\begin{figure}
	\centering
	\resizebox{0.49\textwidth}{!}{%
		\includegraphics{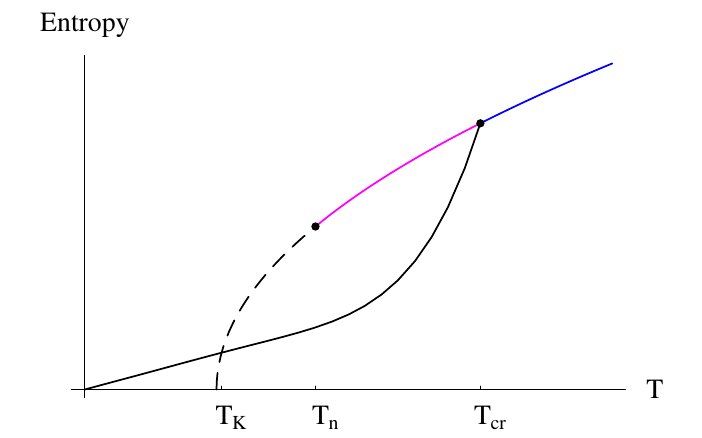}
	}
	\caption{The absolute entropy of the solid, liquid and metastable liquid versus temperature. The black curve represents the entropy of the crystal phase, the blue curve represents the entropy of the liquid phase, the magenta curve represents the entropy of the metastable liquid phase and the dashed curve is the Kauzmann extension of the entropy of the metastable liquid phase. For benzene $T_{\rm cr}$ and $T_{\rm n}$ are given in Table~\ref{Tab1}. In references on the Kauzmann paradox, $T_{\rm n}$ is denoted by $T_{\rm g}$ (the glass temperature). $T_{\rm K}$ is the Kauzmann temperature (intersection of the black and dashed curves).}\label{FigS}
\end{figure}
\begin{figure*}
	\centering
	\resizebox{0.65\textwidth}{!}{%
		\includegraphics{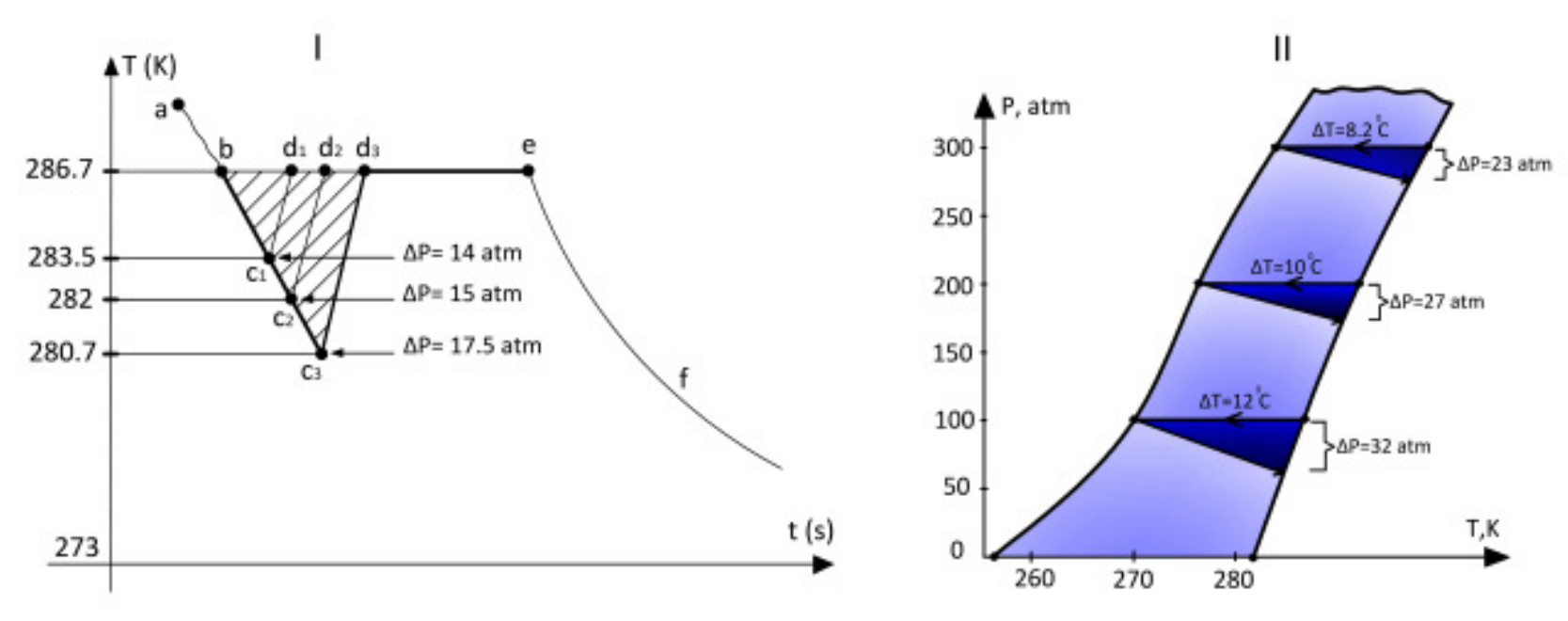}
	}
	\caption{The supercooling  values $\Delta T$ and pressure drop values $\Delta p$, which accompany the crystallization of the benzene metastable state at $p=500$ atm. Plot I: in $T$-$t$ coordinates. Plot II: in $p$-$T$ coordinates.}\label{Fig6}
\end{figure*}

\section{End-point of the metastable state\label{secep}}
In our experiments, the deaerated liquid benzene completely filled the \emph{rigid} container. The \emph{isochoric} cooling resulted in a pressure drop $\Delta p$ that is roughly related to $\Delta T$ by
\begin{equation}\label{ep1}
	\Delta T\approx \frac{\kappa_T}{\alpha}~\Delta p,
\end{equation}
if $\Delta p$ is small. Here $\kappa_T$ and $\alpha$ are the isothermal compressibility and the thermal expansion coefficient: $V\kappa_T=-(\partial V/\partial p)_T$ and $V\alpha=(\partial V/\partial T)_p$. However, as is well known~\cite{5} and Table~\ref{Tab1} confirms it, the pressure drop is very pronounced in the vicinity of the triple point and around it for an \emph{isochoric} cooling process. In this case, the second $(\Delta p)^2$, and probably the third $(\Delta p)^3$, terms in the series expansion~(\ref{ep1}) are needed. In any case, we have a proportionality relation between $\Delta T$ and $\Delta p$:
\begin{equation}\label{ep2}
	\Delta T\propto \Delta p.
\end{equation}

Let us first note that if one parameter, say $\Delta p$, vanishes at some point on the melting curve in a $p$-$T$ diagram, the other parameter, in this case $\Delta T$, also vanishes and conversely. The vanishing of one or the other parameter at some point on the melting curve may mark the absence of liquid metastability at that point. When this is the case, we label such a point the end-point of metastability. That is, in a $p$-$T$ diagram ($p$ vertical and $T$ horizontal as in Fig.~\ref{Fig4}), the liquid phase ceases to exist on the left of the melting curve beyond the end-point of metastability.

For benzene, upon applying different cooling rates, we were not able to observe a state of metastability, within the sensitivity of our experimental set, for $p\geq 2200$ atm. This empirically signals the existence of an end-point of metastability at $p_{\rm ep}=2200$ atm and $T_{\rm ep}=356$ K (see Table~\ref{Tab1} and Fig.~\ref{Fig3}). This end-point is denoted by $M$ in Fig.~\ref{Fig4}.
\begin{figure*}
	\centering
	\resizebox{0.65\textwidth}{!}{%
		\includegraphics{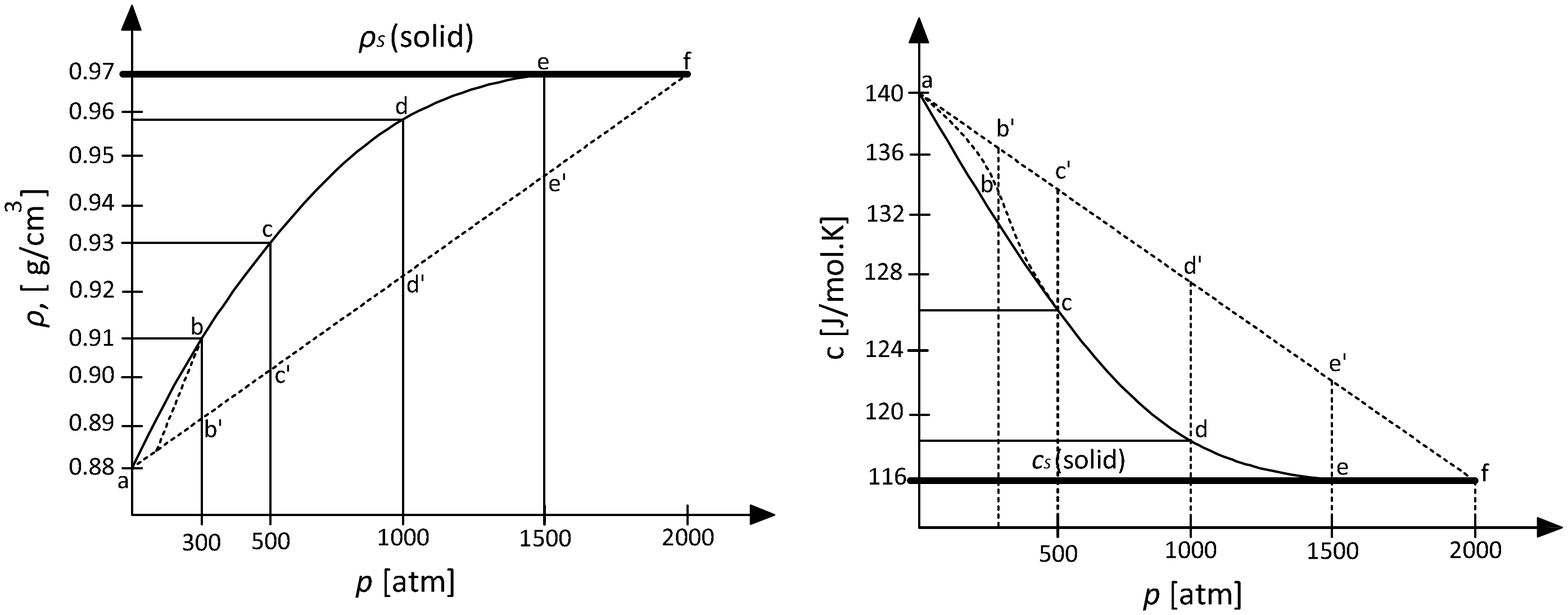}
	}
	\caption{Left Plot: Benzene densities in the metastable region as functions of the pressure. $\rho_{\rm s}$ is the horizontal line at 0.97 g/cm$^3$ and $\rho_{\rm l}$ is the curved line $a\to b\to c\to d\to e\to f$. Right Plot: Benzene molar heat capacities in the metastable region as functions of the pressure. $C_{\rm s}$ is the horizontal line at 116 J/(mol K) and $C_{\rm l}$ is the curved line $a\to b\to c\to d\to e\to f$.}\label{Fig7}
\end{figure*}
\begin{figure}
	\centering
	\resizebox{0.49\textwidth}{!}{%
		\includegraphics{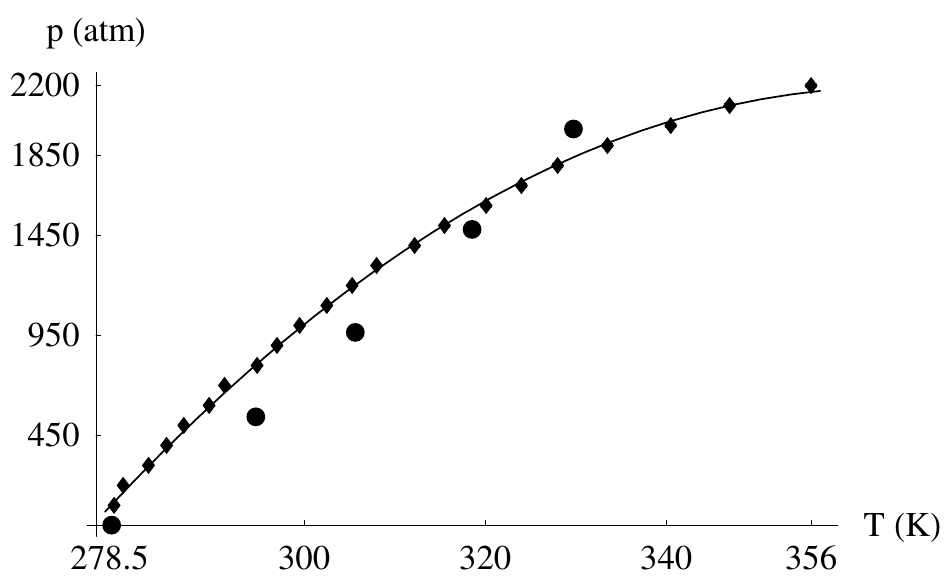}
	}
	\caption{Plot of $p$ versus $T$ using our data shown in Table~\ref{Tab1} (rhombuses). Continuous curve: the parabola $p({\rm atm}) = -35117.6+204.378~T-0.279892~T^2$ ($T$ in K), which provides a good fit to our data points. Recall that the uncertainties are $u_{\rm r}(p) = 0.005$ and $u(T)=0.2$ K. Five large discs: data extracted from Table 2 of Ref.~\cite{data} where the standard uncertainty on the temperature was $u(T)=0.05$ K but the standard uncertainty on the pressure was not given; however we estimated to be $u(p) = 0.1$ MPa.}\label{FigTp}
\end{figure}
\begin{figure*}
	\centering
	\resizebox{0.65\textwidth}{!}{%
		\includegraphics{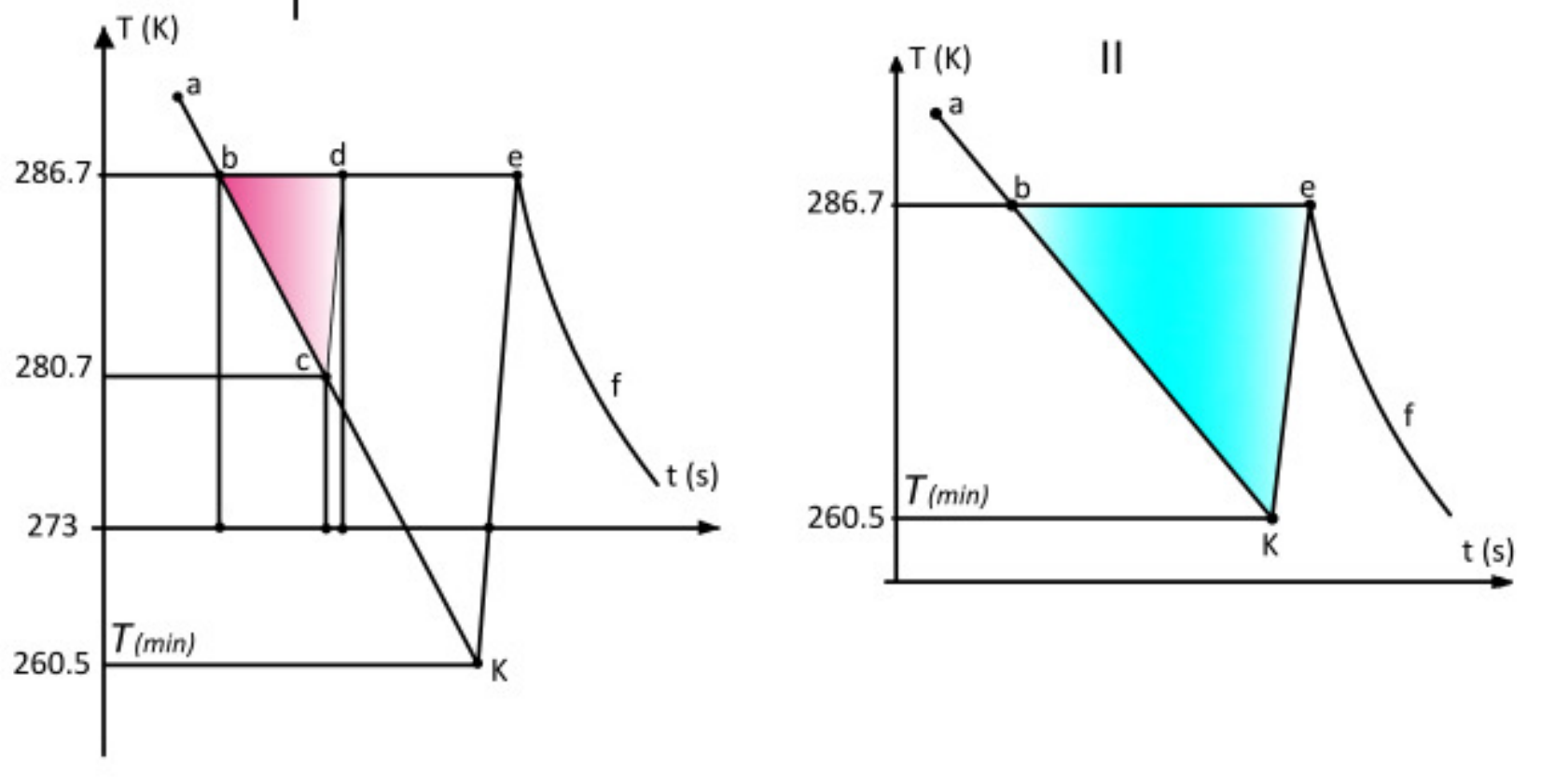}
	}
	\caption{The graphical method for determining the values $\Delta T_{\rm max}$ and  $\Delta p_{\rm max}$ (thermo-gram I) and minimum temperature $T_{\rm min}$  (thermo-gram II) at the pressure of 500 atm.}\label{Fig5}
\end{figure*}

The proportionality of $\Delta V_{\rm m}$ and $\Delta S_{\rm m}$, mentioned earlier in this section, implies that $\Delta S_{\rm m}$ also approaches zero as $p$ approaches 2200 atm. By the Clausius-Clapeyron equation, $dp/dT=\Delta S_{\rm m}/\Delta V_{\rm m}$, this results in an indetermination in the value of the slope $dp/dT$ on the melting line $p(T)$, which can be fixed only empirically, as is the case in the paper by Akella and Kennedy~\cite{AK} where $dp/dT$ assumes a finite value different from zero in the vicinity of 2200 atm (see Fig.~\ref{FigTp} of this paper and Fig.~2 of Ref.~\cite{AK}).

The existence of an end-point of metastability is justified as follows. Figure~\ref{FigS} depicts the entropy versus the temperature. The black curve represents the entropy of the crystal phase, the blue curve represents the entropy of the liquid phase, the magenta curve represents the entropy of the metastable liquid phase and the dashed curve is the Kauzmann extrapolation of the entropy of the metastable liquid phase. For benzene $T_{\rm cr}$ and $T_{\rm n}$ are given in Table~\ref{Tab1}. In references on the {\textquotedblleft Kauzmann paradox\textquotedblright}~\cite{Kauzmann,Stillinger,Cernosek}, $T_{\rm n}$ is denoted by $T_{\rm g}$ (the glass temperature). $T_{\rm K}$ is the Kauzmann temperature (intersection of the black and dashed curves). A liquid phase cannot have an entropy lower than its corresponding crystal or glass phase\footnote{As noticed in Ref.~\cite{Stillinger}, {\textquotedblleft The simple and seemingly reasonable extrapolation of the metastable liquid entropy to temperatures below $T_{\rm g}$ is misleading\textquotedblright}, the Kauzmann extrapolation is justified neither theoretically nor empirically.} (the so-called {\textquotedblleft paradoxical principle\textquotedblright} or {\textquotedblleft Kauzmann paradox\textquotedblright}). Said otherwise, a liquid phase exists only for $T>T_{\rm K}$. Under extreme conditions of high pressure, as is the case with benzene, as the pressure reaches the value of $p_{\rm ep}=2200$ atm, $T_{\rm cr}\to T_{\rm n}\to T_{\rm K}$, that is the three temperature values are nearly equal. Since the liquid phase cannot exist below $T_{\rm K}\simeq T_{\rm cr}$, the metastable state ceases to exist too at the corresponding crystallization temperature of $T_{\rm cr}=T_{\rm ep}=356$ K. We have checked that for $p=2300$ atm there is no metastable state for benzene, however, we do not know if this behavior persists for much higher values of the pressure.

Another point we wish to emphasize is that the existence of an end-point of metastability is not restricted to benzene only. Under extreme conditions of high pressure, this behavior has been noticed in many other materials as shown in the following paragraph. Using the advanced StepScan DSC technique (DSC for Differential Scanning Calorimetry~\cite{Cernosek2,note}), it was shown that the supercooled liquid arsenic triselenide (As\textsubscript{2}Se\textsubscript{3}) may be cooled down to Kauzmann temperature (see Fig~2 of Ref.~\cite{Cernosek}). This indicates that As\textsubscript{2}Se\textsubscript{3} may have its end-point of metastability. For the case of benzene we may claim that for $p<2200$ atm the nucleation temperatures $T_{\rm n}$ given in Table~\ref{Tab1} are much closer, if not equal, to their corresponding Kauzmann temperatures. $T_{\rm n}$ are the lowest values we could reach empirically.

The existence of an end-point of metastability is not characteristic for benzene. We have obtained similar results for meta-toluidine ($p_{\rm ep}=1450$ atm), ortho-toluidine ($p_{\rm ep}=1400$ atm), benzonitrile ($p_{\rm ep}=1275$ atm) and for many other materials that are the subject of our subsequent work. Here the value of the pressure given between parentheses is the pressure at which the metastable liquid phase disappears.

To resume, for benzene the picture we could draw empirically is that the melting curve has, in a $p$-$T$ diagram ($p$ vertical and $T$ horizontal), a left-hand band of metastability of thickness $\Delta T(p)$ that depends on the pressure and extends likely from $p=0$ to $p=2200$ atm. From the triple point up to higher pressures, the thickness of the band as well as the other parameters of the metastable state are inversely proportional to the external pressure.

\section{Treatment of the results - The supercooling $\Delta T$\label{sectr}}
The fact that the temperature  $T_{\rm cr}$  and the temperatures at the points ($d$,  $e$) are increasing functions of the external pressure at equilibrium follows from the famous Clapeyron-Clausius law~\cite{2}. To see why the main parameters of the benzene metastable state, ($t_1$,  $t_3$,  $t_{tot}$), are decreasing function of the external pressure, we should consider the structure of the benzene itself. Based on the charge separation model~\cite{HS}, it was concluded in Ref.~\cite{BG} that benzene liquid has tendency to adhere to the T-shaped structure in which, within the first solvation
shell (a complete shell around a molecule), one molecule lies perpendicular to its nearest one (see Fig.~8 of Ref.~\cite{BG}). We are not concerned with all solid phases of benzene~\cite{Wen}; however, for the range of pressures we are considering in this work (the moderate pressure regime: $p<20$ GPa), the unit cell of solid benzene has the \textit{Pbca} Z = 4 structure~\cite{AC}. As can be seen from Fig.~1 of Ref.~\cite{AC}, this is similar to the T-shaped structure of liquid benzene. As the pressure increases the packing in the liquid benzene mimics that of the orthorhombic phase I (\textit{Pbca}) of the solid phase rendering the liquid-to-solid transition \textquoteleft smooth\textquoteright. This results in a decrease of the degree of supercooling, of the incubation period of a new nucleation phase, and of the time of freezing. As we noticed earlier,  the metastable state disappears at the critical point  $M$ and beyond it. This point is characterized by some critical values of the temperature $356$ K and of the pressure $2200$ atm where any difference between the liquid and crystalline states becomes unnoticeable. As we shall see in Sec.~\ref{seccp}, other studies reached similar conclusions: for instance, it was shown that the molar volume of solid benzene has almost no jump during the solid-to-liquid transition.

A pressure drop $\Delta p$  on the segment $cd$ in the adiabatic process (See Fig.~\ref{Fig3} and Fig.~\ref{Fig6}) may take place, as we have noticed. This effect can be related to changes in the benzene density and heat capacity. The estimations show that the relative values of both densities $\Delta \rho/\rho_{\rm s}$  ($\Delta \rho=\rho_{\rm s}-\rho_{\rm l}$) and of heat capacities $\Delta C/C_{\rm s}$ ($\Delta C=C_{\rm l}-C_{\rm s}$)  are within one and the same order with the relative pressure drop  $\Delta p/p$. A similar picture is observed if we have to do with molar heat capacities. We have noticed that if the pressure is 300 atm, then the ratio $\Delta C/(\delta C_{\rm s})$ is equal to 0.067 [with $C_{\rm s}=116$ J/(mol K), $C_{\rm l}=137$ J/(mol K)~\cite{9}], which is the same as $\Delta p/p$. Here the parameter $\delta$ does not practically depend on $p$ and it is given by
\begin{equation}\label{del}
	\delta\simeq 2.7.
\end{equation}

Taking the above remark into account, we determine the dependence of the liquid density and heat capacity ($\rho_{\rm l},\,C_{\rm l}$) in terms of ($\rho_{\rm s},\,C_{\rm s}$) and $\Delta p/p$ as follows. Notice that ($\rho_{\rm l},\,C_{\rm l}$) may be expressed exactly as
\begin{equation}
\rho_{\rm l}=\rho_{\rm s} \Big(1-\frac{\Delta \rho}{\rho_{\rm s}} \Big),\qquad C_{\rm l}=C_{\rm s} \Big(1+\delta\,\frac{\Delta C}{\delta C_{\rm s}} \Big),
\end{equation}
where no approximation has been made. Admitting that,
\begin{equation}\label{tr3}
	\frac{\Delta p}{p}\simeq \frac{\Delta \rho}{\rho_{\rm s}}\simeq \frac{\Delta C}{\delta C_{\rm s}}\quad {\rm and}\quad \delta \simeq 2.7,
\end{equation}
we obtain
\begin{equation}
\rho_{\rm l}\simeq \rho_{\rm s} \Big(1-\frac{\Delta p}{p} \Big),\qquad C_{\rm l}\simeq C_{\rm s} \Big(1+\delta\,\frac{\Delta p}{p} \Big).
\end{equation}
The values of the parameters obtained from these formulas are given in Table~\ref{Tab2}.

In Fig.~\ref{Fig7} we have depicted plots of ($\rho_{\rm l},\,\rho_{\rm s},\,C_{\rm l},\,C_{\rm s}$) versus the external pressure $p$. As seen from the plots, as the external pressure increases $p$, the supercooled benzene density grows and its molar heat capacity drops from the values corresponding to the liquid state down to the values corresponding to the solid state.  As opposed to linear changes of $\rho_{\rm l}$  and $C_{\rm l}$, the actual changes in the values of these two parameters do not follow the straight path $a\to b'\to c'\to d'\to e'\to f$. Rather, the changes of $\rho_{\rm l}$  and $C_{\rm l}$, as functions of the pressure $p$, take place along the curved path $a\to b\to c\to d\to e\to f$.  For the lower values of $p$, $\rho_{\rm l}$ has a sharp increase ($C_{\rm l}$  has a sharp decrease), and then the shapes of the curves noticeably become smooth as $\rho_{\rm l}$ approaches $\rho_{\rm s}$ (as  $C_{\rm l}$ approaches  $C_{\rm s}$). In these plots $\rho_{\rm s}$ and $C_{\rm s}$ are represented by the horizontal lines at 0.97 g/cm$^3$ (left plot of Fig.~\ref{Fig7}) and at 116 J/(mol K) (right plot of Fig.~\ref{Fig7}), respectively. In both plots there are some deviations of the curves  $\rho_{\rm l}= f (p)$ and  $C_{\rm l}= f (p)$ from the ideal ones but the common tendency of $\rho_{\rm l}$  and $C_{\rm l}$  changes keeps up.
\begin{table*}
	\centering
	\caption{Calculated values of the densities and molar heat capacities of supercooled benzene for different pressure values. Here $\Delta \rho=\rho_{\rm s}-\rho_{\rm l}$ and $\Delta C=C_{\rm l}-C_{\rm s}$ where $\rho_{\rm s}$ and $\rho_{\rm l}$ are the solid and liquid densities of benzene, and $C_{\rm s}$ and $C_{\rm l}$ are its solid and liquid heat capacities at constant pressure. $\Delta p$ is the pressure drop as defined in Table~\ref{Tab1} and $\delta\simeq 2.7$ is a constant parameter~(\ref{del}).}
	\label{Tab2}
	\begin{tabular}{c|c|c|c|c|c|c}
		\hline\noalign{\smallskip}
		$p$   & $\Delta p$    &  $\frac{\Delta p}{p}\simeq \frac{\Delta \rho}{\rho_{\rm s}}\simeq \frac{\Delta c}{\delta c_{\rm s}}$       & $\Delta \rho$           & $\rho_{\rm l}$          & $\Delta c$    & $c_{\rm l}$    \\
		atm    & atm   &  $\delta\simeq 2.7$       & g/cm$^3$    & g/cm$^3$  & J/(mol K)  & J/(mol K) \\
		\hline\noalign{\smallskip}
		1.0  & ---    & ---       & 0.090        & 0.880       & 24.00    & 140.0 \\
		100.0  & 32.0   & ---       & ---           & 0.894       & ---       & ---     \\
		200.0  & 27.0   & ---       & ---           & 0.902       & ---       & ---     \\
		300.0  & 23.0 & 0.0670   & 0.080       & 0.910       & 21.00    & 137.0 \\
		400.0  & 20.0 & 0.0500   & 0.060       & 0.920       & 15.70   & 131.7 \\
		500.0  & 17.5 & 0.0350   & 0.034       & 0.936       & 10.60   & 126.6 \\
		600.0  & 15.0 & 0.0250   & 0.024       & 0.946       & 7.83    & 123.8 \\
		700.0  & 13.0 & 0.0180   & 0.017       & 0.953       & 5.64    & 121.6 \\
		800.0  & 11.5 & 0.0140   & 0.014       & 0.956       & 4.38    & 120.4 \\
		900.0  & 9.5  & 0.0110   & 0.011       & 0.959       & 3.40    & 119.4 \\
		1000.0 & 8.0  & 0.0080   & 0.008       & 0.962       & 2.51    & 118.5 \\
		1100.0 & 6.7  & 0.0060   & 0.006       & 0.964       & 1.88    & 117.8 \\
		1200.0 & 5.5  & 0.0045  & 0.005       & 0.965       & 1.41    & 117.4 \\
		1300.0 & 4.2  & 0.0030   & 0.003       & 0.967       & 0.94    & 116.9 \\
		1400.0 & 3.6  & 0.0020   & 0.002       & 0.968       & 0.60    & 116.6 \\
		1500.0 & 2.8  & 0.0015  & 0.002 & 0.968       & 0.50    & 116.5 \\
		1600.0 & 1.9  & 0.0010  & 0.002 & 0.968       & 0.30    & 116.3 \\
		1700.0 & 1.2  & 0.0007  & 0.001 & 0.969 & 0.20    & 116.2 \\
		1800.0 & 0.6  & 0.0003  & 0.001 & 0.969 & 0.10    & 116.1 \\
		1900.0 & 0.2  & 0.0001  & 0.000  &   0.970 & 0.00 & 116.0 \\
		2000.0 & 0.1  & 0.0000 & 0.000     & 0.970 & 0.00 & 116.0 \\
		2100.0 & 0.0    & 0.0000 & 0.000     & 0.970 & 0.00 & 116.0 \\
		2200.0 & 0.0    & 0.0000 & 0.000     & 0.970 & 0.00 & 116.0 \\
		\hline\noalign{\smallskip}
	\end{tabular}
\end{table*}

Figure~\ref{FigTp} is a plot of $p$ versus $T$ using our data shown in Table~\ref{Tab1} (rhombuses) along with the parabola
\begin{equation}\label{set1}
	p = -35117.6+204.378~T-0.279892~T^2,
\end{equation}
where $p$ is in atm and $T$ in K, which provides a good fit to our data points. We draw a comparison between our data set and the data set given in Table 2 of Ref.~\cite{data} and represented by the five large discs of Fig.~\ref{FigTp}. For the same value of the temperature, the discrepancy between the pressure data, which could reach 257.1 atm at around 294.8 K, is manly due to the use of different experimental sets and to the method itself. Similarly, for the same pressure, the temperature deviations are up to 6.7 K. From this point of view it is worth mentioning that phase transitions are {\textquotedblleft characterized by the hysteresis of melting and crystallization\textquotedblright}~\cite{hysteresis}. This means that the melting temperature is higher than the crystallization temperature with a difference that can reach 400 K~\cite{hysteresis}. Recall that the uncertainties are $u_{\rm r}(p) = 0.005$ and $u(T)=0.2$ K while for the data extracted from Ref.~\cite{data} the standard uncertainty on the temperature was $u(T)=0.05$ K but the standard uncertainty on the pressure was not given; however we estimated to be $u(p) = 0.1$ MPa.

In Ref.~\cite{data} the temperature is in some sense an {\textquoteleft instantaneous\textquoteright} value, that is, the melting {\textquotedblleft temperatures were determined by observing the temperature at which the last trace of solid disappears completely\textquotedblright}~\cite{data}. How is this related to the value of the temperature at the beginnig of the process? In our present work the temperature and pressure were controlled during the whole process of solidification, which lasted up to 426 seconds as shown in Tab.~\ref{Tab1}. So, our data are average values over all fluctuated values. We have been doing such scientific activities repeatedly for years. One of us (B.~\.{I}.) has been working since 1981 on such topics, sometimes under a different name~\cite{F1,F3} (just to mention but a few references), and he has evaluated the phase transitions of benzene and other substances as well~\cite{v1,v2}.

The value of the supercooling $\Delta T$ depends on many factors. In particular, it is known that it increases as the rate of cooling increases~\cite{4}. It is interesting to find out the highest possible supercooling  $\Delta T_{\rm max}$ and the minimum point $T_{\rm min}$ after which the whole volume of liquid benzene freezes at once. To do this, let us apply the graphical method. The thermo-gram of benzene cooling obtained at the external pressure of 500 atm (Fig.~\ref{Fig5}) will be taken as an example.

At the maximum supercooling $\Delta T_{\rm max}$  (or at the minimum temperature $T_{\rm min}$) the adiabatic process of explosive crystallization starts at the point $K$ and the whole sample solidifies at the point $e$. So, one assumes that the temperature increases along the path $K\to e$  in such a way that the line $Ke$ remains parallel to the line $cd$. To determine the location of the point $K$, which is the intersection of the lines $Ke$ and $ac$, we extend the line $ac$ down to the point $K$ in such a way to keep the cooling rate constant (this is achieved upon letting the line $aK$ coincide with the line $cK$). In this case the point  $K$ can be taken as the point corresponding to the minimum temperature  $T_{\rm min}=T_K$ in the region of metastable states and the temperature difference $T_{\rm cr}-T_{K}=\Delta T_{\rm max}$  can be taken as the maximum supercooling. This way we obtained the temperature $T_{\rm min}=T_K=260.5$ K yielding  $\Delta T_{\rm max}= 26.0$ K. As shown earlier, at $p = 0.1$ atm ($\Delta T=20.0$ K) only 27\% of the sample solidifies in the initial stage of crystallization, whereas at 500 atm the whole sample
solidifies (where for this case $T_K\simeq T_{\rm n}$).

Besides the above regularities it was also found out that in the process of adiabatic transitions from the temperature $T_{\rm n}$  to the temperature  $T_d$, minor pressure drops  $\Delta p$  have been determined (see Table~\ref{Tab2} and Fig.~\ref{Fig3}). As the external pressure $p$ increases, the value of the  $\Delta p$ decreases down to zero at the critical point $M$.

For each given value of the external pressure, there is a one-to-one relation between the supercooling $\Delta T$ and the pressure drop $\Delta p$. This phenomenon of overlapping of $\Delta T$ and $\Delta p$ is illustrated in Fig.~\ref{Fig6} I depicting the crystallization of benzene  at the external pressure $p = 500$ atm. We see that $\Delta T$ is an increasing function of $\Delta p$ for $p$ held constant. Table~\ref{Tab1} provides the average values the experimental $\Delta T$ and $\Delta p$ for $p$ held constant. As the external pressure $p$ increases, both $\Delta T$ and $\Delta p$  decrease.

\section{Concluding remarks\label{seccp}}
The near-equality of the densities ($\rho_{\rm l},\,\rho_{\rm s}$), implying the equality of the molar volumes ($V_{\rm m}^{\ \rm l},\,V_{\rm m}^{\ \rm s}$), and of the heat capacities at constant pressure ($C_{\rm l},\,C_{\rm s}$), shown in Table~\ref{Tab2}, which take place in the near vicinity of the end-point $M$ of metastability, are signs that $M$ \emph{might behave} as a critical point for the liquid-solid phase transition since the latter is almost continuous there. The only thing we have not empirically justified here is the equality of the molar entropies ($S_{\rm m}^{\ \rm l},\,S_{\rm m}^{\ \rm s}$) in the near vicinity of the end-point $M$. However, if we rely on the proportionality relation between $\Delta V_{\rm m}$ and $\Delta S_{\rm m}$~\cite{Skripov}, mentioned in Sec.~\ref{secep}, then ($S_{\rm m}^{\ \rm l},\,S_{\rm m}^{\ \rm s}$) are also almost equal at $M$. Despite that, many statements, made in the literature~\cite{Skripov}, do not support the existence of a critical point for the liquid-solid phase transition, claiming that it is impossible for the transition to be continuous due to the structure differences between the solid and liquid phases.

Evidence for a second order pre-melting phase transition in benzene~\cite{Pruzan} constitutes a support to our findings (Table~\ref{Tab2}). This behavior of the solid benzene during melting was first noticed in Ref.~\cite{emph1} and later emphasized in Refs.~\cite{Pruzan,emph2}. It was shown that the molar volume of solid benzene has almost no jump and follows a continuous function until a liquid phase is reached. In this work we, rather, focused on the liquid phase. Our results extend the previous conclusions to the liquid-to-solid transition where, in the near vicinity of the end-point $M$ of metastability, the liquid molar volume and heat capacity approach those of the solid phase (see Table~\ref{Tab2}). These and previous conclusions characterize a two-way nearly second order phase transition. The end-point $M$ of metastability shares some features with the universal liquid-to-gas-transition and gas-to-liquid-transition critical point. The non-shared feature is that beyond the point $M$, while the above-mentioned physical properties (shown in Table~\ref{Tab2}) remain almost undistinguished between the two phases, the two liquid and solid phases are well distinguished as shown in Refs.~\cite{4,Wen,dis1,dis2,dis3,dis4,dis6} and their melting curve extends up to $\sim$5 GPa and $\sim$800 K.

High pressure leads to significant changes in the physical and chemical properties of substances. Benzene has many different crystal structures~\cite{Wen}, it is thus possible that, in the near vicinity of the end-point $M$, one of these crystal configurations is nearly similar (as discussed in Sec.~\ref{sectr}), but with different molar entropy, to the liquid phase due to the high packing of its molecules. This is in some sense the essence of Ostwald's Stage Rule~\cite{rule1} which sates that the phase transition proceeds by steps gradually from a disordered phase (liquid) to less-ordered phase and so on to the most ordered phase (solid). That is, the transition evolves through a series of intermediate metastable phases of increasing stability~\cite{rule2}.

It is worth emphasizing that most workers had examined the solid-to-liquid transition upon fixing the temperature and dropping the pressure. In Ref.~\cite{8} and in the present work we focused on the liquid-to-solid transition upon fixing the pressure and dropping the temperature. This has allowed us to examine closely the metastability of supercooled benzene. It is this experimental procedure that leads to discover the end-point $M$ of metastability and the liquid-solid shared properties in its vicinity. Such a point was never cited or discussed in the scientific literature.

\section*{Acknowledgment} We thank F.~Yal\c{c}{\i}nkaya for providing data graphics support.

\section*{Author contribution statement}
B\.{I} carried out the experimental tests and obtained Table~\ref{Tab1}. MA carried out all the theoretical work and results and edited the manuscript. All authors have read and approved the final manuscript.
%
%

\end{document}